\documentclass[prb,aps,twocolumn,draft,tightenlines,
showpacs,floatfix,superscriptaddress]{revtex4}
\usepackage{psfig}
\usepackage{amssymb}
\usepackage{amsmath}
\usepackage{colordvi}

\begin{document}

\title{Intense terahertz laser fields on a quantum dot with Rashba
  spin-orbit coupling}
\author{J. H. Jiang}
\affiliation{Hefei National Laboratory for Physical Sciences at
  Microscale, University of Science and Technology of China, Hefei,
  Anhui, 230026, China}
\author{M. Q. Weng\footnote{Electric address: mweng@stevens.edu}}
\affiliation{Department of Physics and Engineering Physics,
Stevens Institute of
  Technology, Castle Point on Hudson, Hoboken, NJ 07030, USA
}
\author{M. W. Wu\footnote{Electric address: mwwu@ustc.edu.cn}}
\affiliation{Hefei National Laboratory for Physical Sciences at
  Microscale, University of Science and Technology of China, Hefei,
  Anhui, 230026, China}
\affiliation{Department of Physics,
  University of Science and Technology of China, Hefei,
  Anhui, 230026, China\footnote{Mailing Address}}

\date{\today}
\begin{abstract}
  We investigate the effects of the intense terahertz laser field and
the spin-orbit coupling on  single electron spin in a quantum dot.
The laser field and the spin-orbit coupling can strongly affect
the electron density of states and can excite a magnetic moment.
 The direction of the magnetic moment depends
 on the symmetries of the system, and its amplitude
  can be tuned by the strength and frequency of the
  laser field as well as the  spin-orbit coupling.
\end{abstract}

\maketitle

The study of semiconductor based opto-electronics, which combines
the ultrafast electronics with the low-power optics, has been fruitfully
carried out for years. Schemes based on semiconductor quantum dots
(QD's) have been proposed and actively studied in order to realize the
more ambitious goal of quantum opto-electronic devices, which provide
interface between quantum bits (qubits) and single-photon quantum
optics and can perform quantum information processing and communication on one
chip. Single spins in QD's are natural
candidates\cite{Loss1,Loss2,Loss3} for qubits as they have long
coherence times and can be manipulated both electronically and
optically. There are many theoretical works on spin
relaxation/decoherence and manipulation in
QD's.\cite{LossDot,spintronics,das,wang} Experimental realization of
the electrical/optical generation and read-out of single spin in QD
have also been carried out.\cite{readout1,readout2}

The spin-orbit coupling (SOC) plays an essential role in the dynamics of
electron spins in QD's. It enables the electrical manipulation by
tuning the gate voltage but also leads to the spin relaxation/decoherence
together with the different scattering mechanism such as the electron-phonon
scattering.\cite{LossDot} In addition to the
above effects, the SOC in QD's causes a spin splitting of a few meV,
the order of terahertz
(THz), and therefore is expected to
affect the response of electrons in QD's to the THz laser.
The effect of the SOC on far-infrared optical absorption spectrum has
been discussed in
detail by Rodriguez {\em et al.}\cite{Rodriguez}  Until recently, the
study on the response to the THz radiation has been focused on the
spin-unrelated problems such as the dynamic Franz-Keldysh effect
(DFKE),\cite{Yacoby,Jauho,Jauho2,Nordstrom}
the AC Stark effect\cite{Nordstrom,AH} and the photo-induced
side-band effect.\cite{Cerne,Kono,Phillips,PR1,PR2,Arrachea}
Very recently Cheng and Wu  discussed the effects of intense THz laser
field on a two dimensional electron gas (2DEG)  with the Rashba
SOC.\cite{cheng_2005_APL} It is demonstrated that the laser field can
significantly modify the density of states (DOS) of the 2DEG and
induce a finite off-diagonal density of spin polarization which indicates
the {\em  strong} correlation of different spin branches. Later Wang
{\em et al.} discussed the time-dependent spin-Hall current in the
2DEG driven by an intense THz field.\cite{Wang_Lei} How the
intense THz laser affects the QD's
in the presence of the SOC needs to be
further revealed.
In this paper, we study the effect of intense THz laser
on semiconductor QD's with the Rashba SOC.\cite{Rashba}

We consider a quantum dot grown in an InAs quantum well
with growth direction along the $z$-axis. A uniform
radiation field (RF) ${\mathbf E}(t) = {\mathbf E}\cos(\Omega t)$ is applied
along the $x$-axis with the period $T_0=\frac{2\pi}{\Omega}$. By using
the Coulomb gauge, the Hamiltonian is written as\cite{Winkler}
\begin{equation}
H(t) = \frac{{\mathbf P}^2}{2 m^{\ast}} + V_c({\mathbf r}) +
H_{so}(\mathbf P). \label{eq:Hamiltonian}
\end{equation}
Here the momentum ${\mathbf P} = -i\mbox{\boldmath$\nabla$\unboldmath} +
e{\mathbf A}(t)$ with $\mathbf{A}(t)=-\mathbf{E}\sin(\Omega t)/\Omega$
being the vector potential; $m^{\ast}$ is the
electron effective mass.
The confining potential of the quantum dot is taken to be
$V_c({\mathbf r}) = \frac{1}{2} m^{\ast} \omega_0^2 r^2$
in the $x$-$y$ plane. An infinite-well-depth assumption is made along the
$z$-axis and the well width is assumed to be small
enough so that only the lowest subband is relevant.
$H_{so}$ is the SOC which is composed of the Rashba term\cite{Rashba} and
the Dresselhaus term.\cite{Dresselhaus}
For InAs quantum well, the Rashba term is the dominant one.
$H_{so}({\mathbf P, t})=\alpha_R[\hat{\sigma}_x {P}_y -
\hat{\sigma}_y {P}_x]$ with
\boldmath$\hat{\sigma}$\unboldmath\ denoting the Pauli matrix and
$\alpha_R$ representing the Rashba SOC parameter which can be tuned by
gate voltage up to the order of 4$\times 10^{-9}$\
eV$\cdot$cm.\cite{Grundler,Sato}

By employing the  Floquet theorem, the solution to the
Schr\"odinger equation with time-dependent Hamiltonian (\ref{eq:Hamiltonian})
can be written as,\cite{shirley}
\begin{eqnarray}
  \Psi_{\lambda}({\bf r}, t) &=& e^{-i[E_{em} t-\gamma\sin(2\Omega
    t)]}\nonumber\\
  &\times& e^{-i\varepsilon_{\lambda} t}
  \sum_{n=-\infty}^{\infty}\sum_{\alpha}F_{n\alpha}^{\lambda}
  \phi_{\alpha}({\bf r}) e^{in\Omega t}\ .
\end{eqnarray}
Here $E_{em}=\frac{e^2E^2}{4m^{\ast}\Omega^2}$ is the energy induced by the
RF due to the DFKE;\cite{Jauho,Jauho2} $\gamma = E_{em} / 2\Omega$;
$\{\phi_{\alpha}({\mathbf r})\}$ can be any set of complete basis.
$\{F_{n\alpha}^\lambda\}$ are the eigenvectors of the  equations:
\cite{shirley}
\begin{eqnarray}
  \sum_{m=-\infty}^{\infty}\sum_{\beta} \langle\alpha n|{\cal H_F}|\beta m\rangle
  F_{m\beta}^{\lambda} = \varepsilon_{\lambda} F_{n\alpha}^{\lambda}\ ,
  \label{eigen}
\end{eqnarray}
in which  $\langle{\bf r},t|\alpha n\rangle\equiv\phi_{\alpha}({\bf r}) e^{in\Omega t}$,
${\cal H_F} = H(t) - i\partial_t$ and $\langle\alpha n|{\cal H_F}|\beta m\rangle = H_{\alpha
  \beta}^{n-m} + m\Omega \delta_{\alpha \beta}\delta_{n m}$
with $H^n = \frac{1}{T_0} \int_{0}^{T_0}\!\!d{t}
e^{-in\Omega t} H(t)$ representing the $n$-th Fourier component of the Hamiltonian.
 The eigen-values $\varepsilon_{\lambda}$ of the equations are
solved in the region $(-\Omega/2,\Omega/2]$.\cite{shirley}
In the case of the Rashba SOC,  the non-zero terms are
\begin{eqnarray}
H^0 &=& \frac{{\mathbf p}^2}{2 m^{\ast}} + \frac{1}{2} m^{\ast}
\omega_0^2 r^2 + \alpha_R[\hat{\sigma}_x p_y - \hat{\sigma}_y p_x],
\label{eq:H0}
\\
H^1& =&(H^{-1})^\dagger= \frac{ieE}{2m\Omega}p_x -
\frac{i\alpha_ReE}{2\Omega} \hat{\sigma}_y,
\label{eq:H1}
\end{eqnarray}
with ${\mathbf p} =  -i\mbox{\boldmath$\nabla$\unboldmath}$.

We choose the complete basis $\{\phi_{\alpha}\}$ to be the eigenstates
of the single spin in a QD without the RF and the SOC, whose
Hamiltonian  is described by the first two terms of
Eq.\ (\ref{eq:H0}). These states are identified as
$\{|N,l,\sigma\rangle\}$ with
\begin{eqnarray}
  \langle{\mathbf r}|N,l,\sigma\rangle &=& \sqrt{\frac{n_r!}{a^2\pi
      (n_r+|l|)!}} \nonumber\\
&&\hspace{-0.5cm}\times (r/a)^{|l|} e^{-(r/a)^2/2} {\mathbf
    L}^{|l|}_{n_r}((r/a)^2)e^{il\theta} \chi_{\sigma}\ .
\label{eq:wavef}
\end{eqnarray}
Here ${\mathbf r} = (r\cos\theta, r\sin\theta)$; $a =
\sqrt{\hbar/m^{\ast}\omega_0}$;  ${\mathbf L}^{|l|}_{n_r}$ is the
Laguerre polynomial; $n_r=(N-|l|)/2$; and $\chi_{\sigma}$ is the eigenspinor of
$\sigma_z$. With these basis functions
 one is able to write out the matrix ${\cal H_F}$
and numerically diagonalize  it.\cite{cheng_dot} In this way, one
obtains the eigenvalues $\varepsilon_{\lambda}$ and the eigenvectors
$\{F_{n\alpha}^{\lambda}\}$.

Without the RF and the SOC,
the eigenstates of a single spin in the QD
$|N,l,\sigma\rangle$ are $2\times(N+1)$-fold degenerate in angular momentum
(index $l=-N,-N+2,\cdots,N-2,N$) and spin momentum (index $\sigma=\uparrow$,
$\downarrow$),
 with energy being $(N+1) \omega_0$. The
degeneracy is partially lifted by the SOC into $(N+1)$ states,
each with 2-fold Kramers degeneracy. Although these states are no
longer eigenstates of $\sigma_z$, they can be distinguished by the
corresponding majority spin components as quasi--spin-up and -down
states.\cite{Rodriguez,cheng_dot} The schematic of the lowest 12
states  together with the possible transitions
among them is shown in Fig.\ 1. For both the small
SOC and the small RF, the transitions are mainly between the states
in the same quasi-spin branch (the solid arrows in Fig.\ 1),
since the RF does not flip the spin. At
high RF and/or high SOC, the direct transitions (the dotted arrows)
and multi-photon processes between different quasi-spin
branches are important, therefore the two quasi-spin branches
become strongly mixed.
The mixing can be roughly quantified by the quantity $\Lambda =
\alpha_R eE/(\Omega^2)$ which comes from the photon-assisted spin
flip (the second term of $H^1$) divided by the THz
frequency.\cite{cheng_2005_APL}
The stronger the correlation between different spin branches
is, the larger $\Lambda$ becomes.

\begin{figure}[htb]
  \centerline{
    \psfig{figure=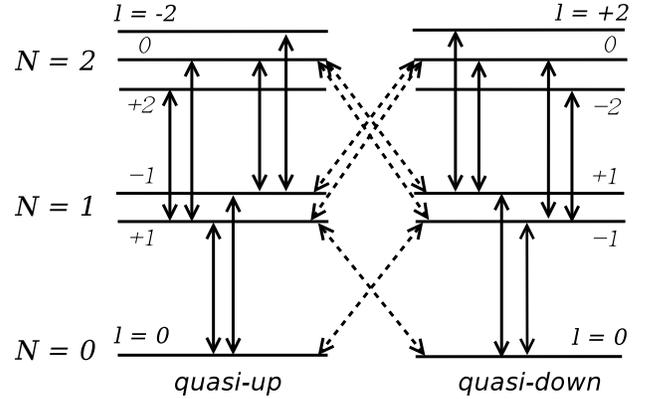,width=0.95\columnwidth}
  }
  \caption{Schematic of the  lowest 12 states of zero field
    Hamiltonian $H^0$, with all possible  transitions.
    The transitions are indicated by solid arrows for
    quasi-spin conserving transitions and dotted arrows for
    quasi-spin flip transitions.
  }
\label{fig:state}
\end{figure}

The effects of the RF and SOC can be qualified by the
density of states (DOS), $\rho_{\sigma,\sigma}$,
and density of spin polarization (DOSP),
$\rho_{\sigma,-\sigma}$.\cite{cheng_2005_APL}
These densities can be calculated from spectral functions which can
be extracted from the Green functions:
\begin{widetext}
\begin{eqnarray}
 && {\mathbf \rho}_{\sigma_1,\sigma_2}(T, \omega)
 = -\frac{1}{2\pi i} \int\limits_{-\infty}^{\infty}d{t} e^{i\omega t}
  \int d^{2}\mathbf{r}
 \bigl\{
G^{r}_{\sigma_1\sigma_2}(\mathbf{r},t_1,t_2)-
G^{a}_{\sigma_1\sigma_2}(\mathbf{r},t_1,t_2)
 \bigr\}
  \nonumber\\
  && = \frac{1}{2\pi} \int\limits_{-\infty}^{\infty}\!\!\!d{t}
  e^{i\omega t} \iint\limits_{-\infty}^{\ \ \infty}\!\!\!d{\bf
    r} \sum_{\lambda} \Psi_{\lambda}^{\sigma_1}({\bf
    r},t_1) {\Psi_{\lambda}^{\sigma_2}}^{\ast}({\bf r},t_2)\nonumber\\
  & &= \sum_{p,n,m,\lambda,\alpha,\beta}
  g(\sigma_1,\sigma_2;\alpha,\beta)F_{n\alpha}^{\lambda}{F_{m\beta}^{\lambda}}^{\ast}  
J_{p}(2\gamma\cos(2\Omega T))\exp[i(n-m)\Omega T] \nonumber\\
&&\hspace{0.5cm}\times
\delta(\omega-[\varepsilon_{\lambda}+E_{em}-(2p+n+m)\Omega/2]),
\label{rho}
\end{eqnarray}
\end{widetext}
where $G^{r}$ and $G^{a}$ are retarded and advanced Green's
  functions respectively, $T=(t_1+t_2)/2$, $t=t_1-t_2$,
$\alpha$ and $\beta$ stand for $\{N,l,\sigma\}$, $J_p$ is  the Bessel
function of  $p$-th order and
\begin{eqnarray}
&&g\bigl(\sigma_1,\sigma_2;\alpha=\{N,l,\sigma\},
\beta=\{N^{\prime},l^{\prime},\sigma^{\prime}\}\bigr)\nonumber\\
&& =\langle \phi^{\sigma_1}_{\alpha}|
\phi^{\sigma_2}_{\beta}\rangle
=\delta_{N,N^{\prime}}\delta_{l,l^{\prime}}\delta_{\sigma_1,\sigma}
\delta_{\sigma_2,\sigma^{\prime}}\ .
  \label{eq:g}
\end{eqnarray}
It is seen from Eq.\ (\ref{rho}) that these densities are periodic
functions of $T$ with period $T_0$. Symmetry analysis (time
reversal symmetry and parity symmetry) further suggests that there
is no spin polarization along the $x$- and $z$-axis. The only
non-vanishing component of the spin polarization is along the
$y$-axis and reduces to zero if the RF is turned off. The
spin polarization is an odd function of time
$\rho_{\sigma,-\sigma}(T,\omega)=-\rho_{\sigma,-\sigma}(-T,\omega)$,
and therefore there is no overall spin polarization once averaged
over time. On the other hand, the DOS is an even
function of time and
$\rho_{\uparrow,\uparrow}(T,\omega)=\rho_{\downarrow,\downarrow}(T,\omega)$.
Thus the RF and the SOC can affect the overall DOS, but cannot
induce any spin polarization along the $z$-axis. In order to
account for the effect of the RF and the SOC on the system, we use the average
of $\rho_{\sigma\sigma}(T,\omega)$ over a period to
quantify the DOS
\begin{equation}
  \label{eq:DOS}
  \bar{\rho}_{\sigma\sigma}(\omega)={1\over T_0} \int_0^{T_0}
  \rho_{\sigma,\sigma}(T,\omega)\, dT.
\end{equation}

By numerically diagonalizing the matrix ${\cal H_F}$, one is able to
obtain the coefficients $\{F^{\lambda}_{n\alpha}\}$ and then
calculate the DOS and the DOSP through Eq.\ (\ref{rho}). One can further
obtain the time-averaged DOS by using Eq.\ (\ref{eq:DOS}) and
calculate the magnetic moment along the $y$-axis $M_y$ via proper evaluation
  of $\langle \sigma_y\rangle$.\cite{cheng_2005_APL}
To converge the DOS in the energy range between 0 to
$2.4\omega_0$, one has to use as
many as 132 states
from the lowest 11 major shells ($N=0$-$10$)
since the energy levels of the
QD are almost equidistant and hence lead to significant resonant
overlap.\cite{shirley}
In the following we present the numerical results of
the time-averaged DOS and the magnetic moment $M_y$
under different conditions.
In the calculation, the effective mass
$m^{\ast}=0.0239m_0$ with $m_0$ representing
the free electron mass and the Rashba coefficient $\alpha_R=
3\times 10^{-9}$\ eV $\cdot$ cm.\cite{Grundler,Sato}
 The confinement potential $\omega_0$ is chosen to be 5\ THz, which
corresponds to a QD with diameter about
 $55$\ nm.\cite{cheng_dot,LossDot,Hanson}

\begin{figure}[htb]
\centerline{\psfig{file=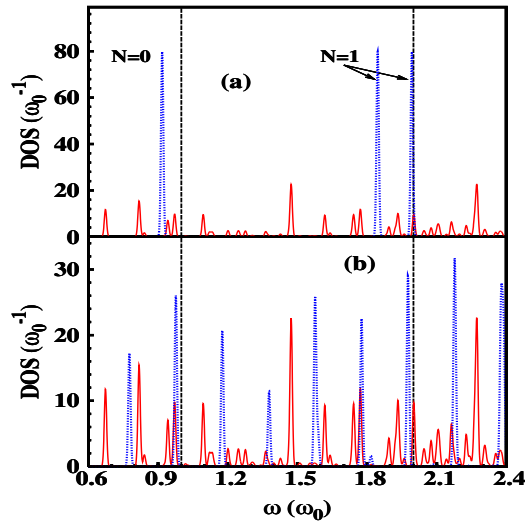,width=7cm,height=7cm}}
\caption{(Color online) (a) The time-averaged DOS of QD under THz field with
    $E=0.6$\ kV/cm, $\Omega=4$ THz (red solid curve) and $E=0$ (blue dotted curve)
    for $\alpha_R=3\times 10^{-9}$\ eV$\cdot$cm. (b) The
    time-averaged DOS of QD with Rashba SOC $\alpha_R=3\times
    10^{-9}$\ eV$\cdot$cm (red solid curve) and $\alpha_R=0$ (blue dotted curve)
    for $E=0.6$\ kV/cm, $\Omega=4$\ THz.
 The black dashed lines correspond to the
    positions of $\omega_0$ and $2\omega_0$.}
  \label{fig:DOS1}
\end{figure}

In Fig.\ 2  we compare the time-averaged DOS
with and without the SOC or the RF. The red (solid) curves are the DOS with
both the SOC and the RF when $E=0.6$\ kV/cm and
$\Omega=4$\ THz. The blue (dotted) curve in Fig.\ 2(a) is the DOS
with only the SOC while the one in (b) is the DOS with only the RF. It is noted that
in the figure the $\delta$-function $\delta(x)$ in the DOS
is replaced by the Gaussian function
$\exp(-x^2/2\sigma^2)/(\sigma\sqrt{2\pi})$ with $\sigma=0.005\omega_0$.
The DOS  without the SOC and the RF
peaks at $\omega=(N+1)\omega_0$ with value $(N+1)/(\sigma\sqrt{2\pi})
(N=0,1,2,\cdots)$ (with the lowest two levels labelled by dashed lines
in the figure). Each level has $(N+1)$-fold degeneracy
as  pointed out above. Nevertheless,
by including only the SOC, the 2-fold degeneracy of $N=1$ major shell is lifted
and the ground state energy ($N=0$) is lowered by $0.1\omega_0$, as
indicated in Fig.\ 2(a).
Moreover, including only the RF gives rise to many peaks at the
integer multiplications of $\Omega$, as can be seen
in Eq.\ (\ref{rho}).  It is seen from Fig.\ 2(b) that
these peaks are equidistant.
The peaks are generally higher than those in the case with both the
RF and the SOC, as the SOC causes additional splitting of the
degenerate states.

\begin{figure}[htb]
\centerline{\psfig{file=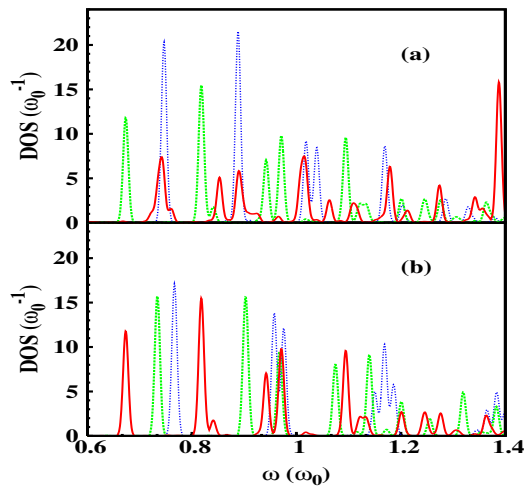,width=7cm,height=6.5cm}}
\caption{(Color online) The time-averaged DOS (a)
    for $E=0.5$ (green dot-dashed curve),
    0.6 (blue dotted curve) and 0.7 (red solid curve)\ kV/cm,
    with $\alpha_R=3\times 10^{-9}$\ eV$\cdot$cm and $\Omega=4$\ THz
    and (b) for $\alpha_R=1$ (green dot-dashed curve),
 2 (blue dotted curve) and 3 (red solid curve) $\times 10^{-9}$\ eV$\cdot$cm,
    with $E=0.6$\ kV/cm and $\Omega=4$\ THz.}
\label{fig:RF}
\end{figure}

In order to further reveal the effect of the SOC and the RF on the system, we
study the DOS under different RF and SOC. In Fig.\ 3, the
DOS with different
RF strengths (a) and  Rashba coefficients (b)
are plotted against energy.
It is seen from Fig.\ 3(a) that the first peak of the DOS
gets a red shift from $0.75\omega_0$ when $E=0.5$\ kV/cm to
$0.68\omega_0$ when $E=0.6$\ kV/cm
and then comes back to
$0.74\omega_0$ when $E$ is further increased to $0.7$\ kV/cm.
This is a clear demonstration of the combined effect of the AC Stark
effect and the DFKE.\cite{Nordstrom}
The AC Stark effect, which states that the energy difference becomes
larger (smaller) if two energy levels are driven by an AC field with
frequency below (above) the resonance.\cite{Nordstrom,AH} On the other hand,
the DFKE states that the DOS has a blue-shift due to
$E_{em}$ as indicated in Eq.\ (\ref{rho}).\cite{Jauho,Jauho2}
In the current case, the frequency
of the THz irradiation is 4\ THz, which is smaller than the resonant
frequencies of the transitions from $N=0$ to $N=1$ major shells even
when the SOC is
included. Therefore there is an AC Stark effect which shift the
  $N=0$ major shell to red-side.
This effect is responsible for the red shifts
as the DOS in the low energy range mainly consists of $N=0$ major
shell. Although both the DFKE blue-shift $E_{em}$ and the AC Stark
red-shift increase with the RF strength,
the AC Stark effect is more important at low RF strength, and it
saturates at high RF strength where the blue-shift
due to the DFKE dominates.\cite{Nordstrom}
Moreover, at low RF intensity the peaks
locate at $\varepsilon_{\lambda}+E_{em} - n_0\Omega/2$ of a certain $n_0$
or replicas $\varepsilon_{\lambda}+E_{em}-(n_0+n)\Omega/2$ with small
$n$'s. When the RF increases, the multi-photon
processes become more and more
important. As a result, the DOS becomes smoother as more and more
multi-photon replicas appear.
The effect of the SOC  is plotted in Fig.\ 3(b),
where the DOS under different Rashba coefficient $\alpha_R$ is plotted
when the RF $E=0.6$\ kV/cm and
$\Omega=4$\ THz. It is seen that the first peak
shifts to the red side as $\alpha_R$ increases.
This is understood as the SOC contributes to the AC Stark
effect. From Eq.\ (\ref{eq:H1}), one notices that the AC Stark effect
enhances as the SOC is increased. Differing from the
situation shown in Fig.\ 3(a), the blue shift $E_{em}$
from the DFKE does not change since the RF is fixed here.
Therefore the shift of the first
DOS peak is monotonic. Moreover, as $\alpha_R$ increases  the multi-photon
processes between different spin branches
become more important and the DOS becomes smoother.

As said before,  the 
DOSP is not zero when both the SOC and the RF are
present. Due to  the symmetry of the system, the only remaining
spin polarization is along the $y$-axis. Consequently, the induced average magnetic
moment can be calculated through the equation\cite{cheng_2005_APL}
\begin{equation}
  {\bf M}(T)=\left(0,
    g\mu_B
    \int_{-\infty}^{E_F(T)}d\omega\
    \mbox{Im}\rho_{\uparrow,\downarrow}(T, \omega), 0\right)\ ,
  \label{My}
\end{equation}
where the time-dependent Fermi energy $E_F(T)$ is determined by $2
\int_{-\infty}^{E_F(T)}d\omega\ \rho_{\uparrow,\uparrow}(T, \omega)=1$
when there is only one electron in the quantum dot.
Due to the time periodicity introduced by the THz field,
$E_F(T)$ and ${\bf M}(T)$ are also periodic functions of $T$ with
period $T_0$. In Fig.\ 4, we plot the $y$-component of magnetic moment as a
function of time. The magnetic moment is controlled by the RF strength, the frequency
as well as the SOC and it can be qualitatively described by the factor
$\Lambda$. The stronger the RF and the SOC are, the larger $\Lambda$ and $M_y$
become. On the other hand, the larger $\Omega$ is, the smaller $\Lambda$
becomes and therefore the smaller $M_y$ becomes.

\begin{figure}[htb]
\centerline{\psfig{file=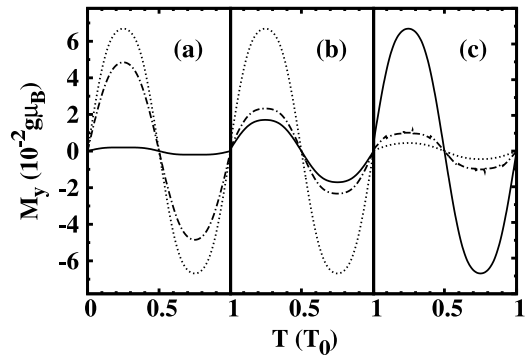,width=0.8\columnwidth}}
  \caption{The average magnetic moment $M_y$ versus the time
    for a single spin in QD at
    (a)
    $E=0.3$ (solid curve), 0.5 (dotted curve), and 0.6
    (dot-dashed curve)\  kV/cm for fixed  $\Omega=4$\ THz,
 and  $\alpha_R=3\times 10^{-9}$\ eV$\cdot$cm;
    (b)
    $\alpha_R=1$ (solid curve), 1.5 (dotted curve), and 3
    (dot-dashed curve)
    $\times 10^{-9}$\ eV$\cdot$cm for fixed RF with $\Omega=4$\
    THz and $E=0.6$\ kV/cm;
    (c) $\Omega=4$ (solid curve), 4.6 (dotted curve), and 6 (dot-dashed
curve)\  THz  for fixed $E=0.6$\ kV/cm and
 $\alpha_R=3\times 10^{-9}$\ eV$\cdot$cm .
}
  \label{fig:My}
\end{figure}

In conclusion, we study the effects of the intense THz field
on single electron spin in QD with the Rashba SOC.
We calculate the single electron DOS at zero
temperature and show that the DOS is greatly affected by the THz field and the SOC
due to the dynamic Franz-Keldysh effect, the AC
Stark effect and the side-band effect. It is shown that
the joint effect of the THz field and the SOC can excite a THz  magnetic
moment which is controlled by the THz field strength, the THz frequency
 as well
as the SOC. This provides a unique way to convert THz electric signals into
THz magnetic ones which may be useful in full electrical  magnetic resonance
measurements.

  This work was supported by the Natural Science Foundation of China
  under Grant Nos. 90303012 and 10574120, the Natural Science Foundation
  of Anhui Province under Grant No. 050460203, the Knowledge Innovation
  Project of Chinese Academy of Sciences and SRFDP.
  J.H.J. would like to thank J. L. Cheng for helpful discussions.

\end{document}